# Effects of Graphene Interface on Potassiation in a Graphene- Selenium Heterostructure Cathode for Potassium-ion Batteries

*Vidushi Sharma\* and Dibakar Datta*

*Department of Mechanical and Industrial Engineering, New Jersey Institute of Technology,*

*Newark, New Jersey 07103, United States*

\*Corresponding Author

Vidushi Sharma, Email: vs574@njit.edu, vidushis@ibm.com

**ABSTRACT**

Selenium (Se) cathodes are an exciting emerging high energy density storage system for Potassium ion batteries (KIB), where potassiation reactions are less understood. Here, we present an atomic-level investigation of $K_xSe$ cathode enclosed in hexagonal lattices of carbon(C) characteristic of layered graphene matrix and multiwalled carbon nanotubes (MW-CNTs). Microstructural changes directed by graphene substrate in $K_xSe$ cathode are contrasted with graphene-free cathode. Graphene's binding affinity for long-chain polyselenides ($Se_3$ = -2.82 eV and $Se_2$ = -2.646 eV) at low K concentration and ability to induce enhanced reactivity between Se and K at high K concentration are investigated. Furthermore, intercalation voltage for graphene enclosed $K_xSe$ cathode reaction intermediates are calculated with $K_2Se$ as the final discharged product. Our results indicate a single-step reaction near a voltage of 1.55 V between K and Se cathode. Findings in the paper suggest that operating at higher voltages (~2V) could result in the formation of reaction intermediates where intercalation/deintercalation of K could be a challenge, and therefore cause irreversible capacity losses in the battery. Primary issue here is the modulating favorability of graphene surface towards discharging of Se cathode due to its differential preferences for K-Se reaction intermediates. A comparison with graphene-free cathode highlights the substantial changes a van der Waals (vdW) graphene interface can bring in atomic-structure and electrochemistry of the $K_xSe$ cathode.

## 1. Introduction

Current innovations in energy sector are focused on balancing the world energy consumption with clean energy solutions. Massive research and development efforts have been made in the past few decades to advance lithium-ion battery (LIB) stature to the primary energy storage for most engineering applications, including electronic devices, transportation, wearables, etc.[1, 2] The rarity of Li and cost associated with development of LIBs strongly advise the use of low-cost complements such as sodium-ion batteries (NIBs) and potassium-ion batteries (KIBs) in applications where maximizing energy and power is not essential such as in small-scale energy applications.[3] Particularly, KIBs have garnered interest as low cost complements of LIBs owing to the relative earthly abundant battery precursor materials. [4-7] Energy storage mechanism in KIB is similar to 'rocking chair' operation in LIBs, except for the respectively larger ion carriers that greatly differentiate the electrode intercalation mechanism and diffusion kinetics in KIB battery systems. This suggests careful tailoring of the structural design of electrodes[3] for next-generation sustainable KIB batteries.

Among low-cost alternatives to Li for energy storage, K has a lower standard reduction potential, which is closer to Li (-3.04V $E^0$ = Li$^+$/Li$^-$ < -2.93V $E^0$ = K$^+$/K$^-$ < -2.71V $E^0$ = Na$^+$/Na$^-$).[3] This permits KIBs to operate at higher potentials with better energy density than NIBs.[8] Moreover, the ionic mobility of K remains unhindered by its weight and is competitive to Li.[9] On the electrode front, the commercial graphite anode used in LIB can be customized easily for KIB. K intercalated graphite is found stable even at a high alkali density of $KC_8$.[10] The electrochemical analysis by Komaba et al. demonstrated graphite anode to have 244 mAhg$^{-1}$ reversible capacity for KIB in 0 - 0.3 V range.[9] Likewise, most commercially acceptable LIB anodes perform well for KIBs at relatively safer voltages.[8] However, KIB cathodes become rather challenging to design due to the large size of K ion. Transition metal layered oxide cathodes exhibit fast capacity fading in KIB because it is not easy to extract or re-intercalate large K ions without any structural damage.[11] Layered birnessite $K_{0.3}MnO_2$ was one of the first layered oxide cathodes investigated for KIB by Vaalma et al.[12] in non-aqueous electrolyte, that only showed a reversible capacity of 65 mAhg$^{-1}$ between 3.5 - 1.5V and 57% capacity retention. Instead, metal-organic frameworks (MOFs) are preferred alternatives to layered oxides as KIB cathodes. Pore sizes on MOF are adjustable and promising for reversible K storage. Eftekhari et al. introduced K containing MOF called Prussian Blue (PB) cathode KFe$_4^{III}$[Fe$^{II}$(CN)$_6$], which could achieve 78.62 mAhg$^{-1}$ reversible capacity and only 12% capacity fade post 500 cycles.[6] This stability marked PB as a prospective cathode for KIB and encouraged experimental electrochemical studies on several PB analogs.[13-16] Despite all efforts, high energy density cathodes still remain a major limiting factor in the development and adoption of KIB.

Chalcogenides like sulfur (S) and selenium (Se) store metal ions by conversion reactions that benefit energy densities. [17-22] However, these promising electrodes suffer from poor reversibility due to dissolution of reaction intermediates in the electrolyte and pulverization caused by large volume expansions.[17-22] Se is comparatively heavier, less reactive, and more electroconductive than S,[23, 24] which imply Se-based cathodes can have good electron transport and better control over shuttle effects if well engineered with confining composite matrix. Liu et al. were the first to report performance of Se cathode confined in carbonized polyacrylonitrile for KIB in 2017.[25] Active Se in composite maintained a reversible capacity of 396 mAhg$^{-1}$ with $K_2Se$ as a final discharged product. One added problem identified for K-Se batteries is the limited reactivity of both metals due to their large atomic size.[26] The strategy of confining Se in carbon (C) allotropes to control surface reactivity, phase transition, and volume change in Se cathode has been well explored for LIBs and next-generation metal batteries.[20, 21, 27-36] However, the role of Se supporting C matrix goes beyond polyselenide confinement to providing an interface for easy expansion and contraction of Se during battery cycles. [37, 38] Interestingly, refined hexagonal C lattice such as graphene(Gr) have the advantage of a slippery van der Walls surface that is shown to be effective in combating stresses in alloying electrodes upon ion storage, thereby improving the cycle life of the composite electrode.[37] We previously reported the detailed characterization of the Se-Gr interface and its promising potential in mitigating interfacial stresses due to low interface adhesion (0.43 J/m$^2$)[38], which was comparable to the silicon (Si) – Gr system reported by Basu et al.[37] As such, using graphene-based supporting matrix for Se cathode can achieve major electrode design targets like high electronic conductivity, physical confinement, large surface area, alleviated volumetric and mechanical stresses.[38] However, the involvement of graphene in electrochemistry of K-Se battery is still less understood in literature.

In present work, we further explore the effects of interface presented by graphene to Se cathode in a heterostructure electrode for application in next-generation KIB battery by computational methods. Latest simulation results have confirmed that hexagonal lattice of C has a strong chemical affinity for $K_2Se$.[27] Therefore, we use $K_2Se$ as the final discharged product of Se and compare the electrochemical voltage profile of graphene-free Se cathode with graphene-supported Se cathode. K intercalation and distribution are modeled in reaction intermediates between Se cathode and discharged product $K_2Se$ for graphene-free as well as graphene-supported configurations. Microstructural changes at graphene- $K_xSe$ interface are investigated and further decomposed into Se-K clusters, whose binding energy with graphene substrate is calculated. Next, we throw some light on influence of graphene interface on electrochemical mechanism in Se-K cathode. We calculate the operational voltage to direct single-step $K_2Se$ cathode discharging in a graphene enclosure. Substantial electrochemical changes that a graphene interface brings in the cathode without any covalent bonding with Se-K have been highlighted.

## 2. Methodology

### 2.1. Structure of Graphene supported Se cathode

Two-dimensional (2D) materials such as graphene can form heterostructure electrodes with active materials by multiple design strategies shown in Figure 1(a) including mixed, wrapped, encapsulated, and layered approach. Though the exact chemical steps for experimentally preparing these electrode designs can deviate, the general approach involves infusing Se in graphene-based matrix at high temperature conditions followed by condensation.[39, 40] Sha et al. reported synthesis of layer-by-layer stacked Se-graphene cathode by high energy ball milling procedure.[41] During the process, graphene gets fragmented while crystalline Se turns amorphous and distributes across the graphene surface. These variations in macroscopic designs of the Se-Gr heterostructures majorly influence K diffusion pathways, mechanical stability, and electrolyte - electrode interface stability. However, if we zoom into the atomic scale (Figure 1(b)), the electrochemical properties of the interface formed between Se cathode and graphene will stay consistent across the design space. Our study presents atomic-level investigation of Se-Gr composites/ heterostructures as cathodes.

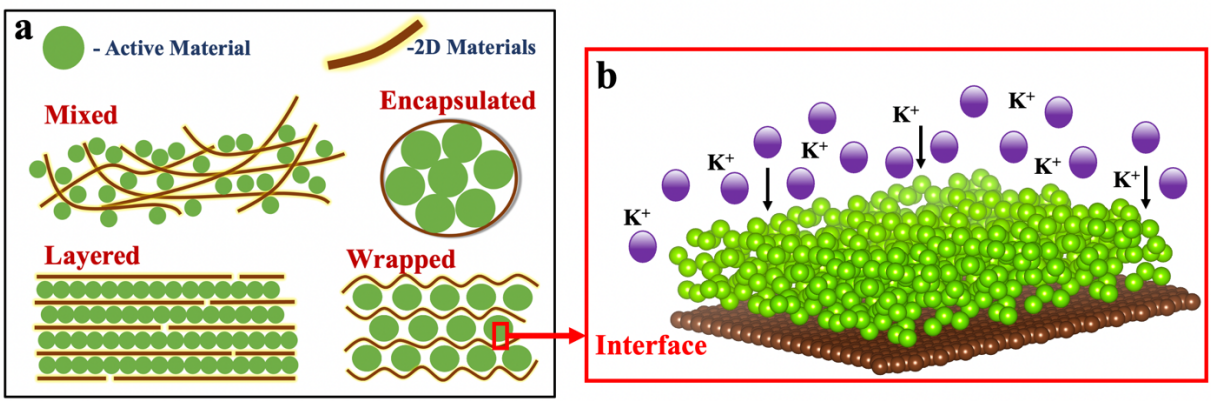

**Figure 1. (a)** Schematic representation of electrode design strategies with active material and two-dimensional (2D) materials such as graphene. **(b)** Atomic representation of the interface formed between active material Se and graphene in various electrode designs.

Amorphous Se (a-Se) was derived by computational quenching of monoclinic selenium-containing $Se_8$ rings that break to form different sized polymeric chains with Se-Se bond length of 2.4 Å. The complete computational modeling and characterization of a-Se and a-Se/Gr have been reported in our previous study.[38] In battery systems, Se is mostly present in amorphous form with $Se_n$ chains (n = 2 to 8).[36] Alternatively, if originally used in crystalline form, $Se_8$ rings are converted to $Se_n$ chains after the first battery cycle and remain so for the rest of the battery life.[42]

Structural parameters such as interatomic bond lengths, bond angles, and dihedral angles are comparable among pristine Se allotropes,[43, 44] yet graphene matrix surface can direct the aligned distribution of Se.

To generate 3D periodic configurations of a-$K_x$Se cathode supported over graphene substrate, we started with an optimized amorphous a-$Se_{30}$ system (Figure 2(a)) and sequentially added 6 K atoms at a time until a-$K_{60}Se_{30}$ (a-$K_2$Se) is achieved (Figure 2(b)). After each potassium addition step, K atoms were allowed to diffuse in the cathode during an ab initio molecular dynamics simulation (AIMD) run and then relaxed with density functional theory (DFT) until energy-optimized a-$K_x$Se structures were obtained. Volume of the simulation cell was allowed to ease in all dimensions. Between initial Se and final a-$K_{60}Se_{30}$ (Figure 2(a-b)), 183.52% volume expansion was noted. The simulation cell's base also expanded during potassiation, and its x-y dimensions were used to determine graphene substrate's size. The final a-$K_{60}Se_{30}$ bulk configuration was placed on top of a periodic graphene lattice containing 96 $sp^2$ hybridized carbon atoms with the interfacial gap of ~2.8 Å to form the a-$K_{60}Se_{30}$/Gr interface (a-$K_2$Se/Gr). Note that the surface area of graphene substrate was equivalent to the x - y surface of final a-$K_{60}Se_{30}$. We further let the atoms diffuse and re-adjust on the graphene lattice during an AIMD run followed by DFT optimization. For graphene-supported K-Se cathode, optimized a-$K_2$Se/Gr configuration was considered the final discharged product (Figure 2(c)). Starting from this a-$K_2$Se/Gr structure, 6 K atoms were sequentially removed, followed by AIMD run and complete energy optimization, until a-$Se_{30}$/Gr is left (Figure 2(d)). Removal of K atom from the system at each step was completely random and system atoms were allowed to diffuse to find equilibria during the following AIMD run and optimization. Upon complete charging (depotassiation), the end structure resembles a-Se clusters distributed on a periodic graphene mesh (Figure 2(a)). This computationally modeled configuration is close to experimentally synthesized nano architectures of active electrodes and graphene.[39, 41, 45, 46]

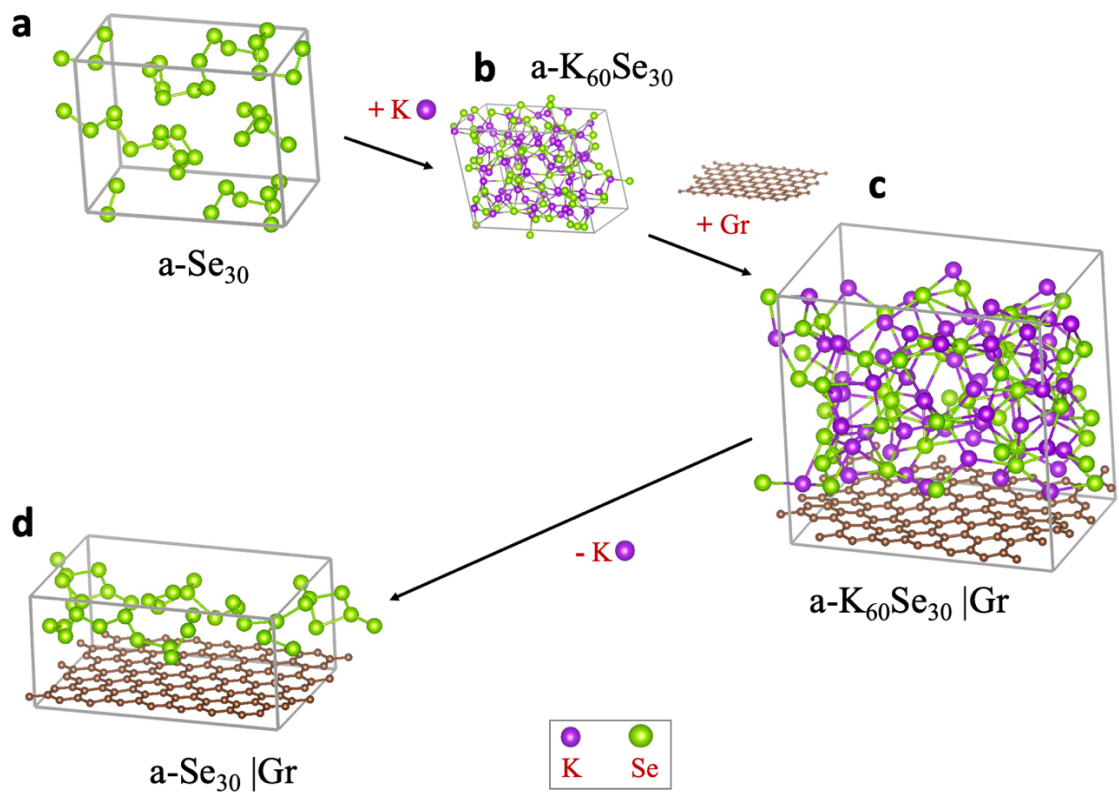

**Figure 2. (a)** Atomic representation of initial optimized amorphous Se (a-Se$_{30}$) generated from a crystalline Se with computational quenching. The structure is dominated by disintegrated forms of Se rings as chains with nearest neighboring distance of ~2.4Å. **(b)** Miniaturized view of a-K$_{60}$Se$_{30}$ (a-K$_2$Se) generated after complete potassiation (discharging). The volume is expanded by 183.52 %. **(c)** Atomic representation of a-K$_{60}$Se$_{30}$ /Gr (a-K$_2$Se/Gr) system. Graphene with surface area equivalent to the base of *a*-K$_{60}$Se$_{30}$ was added in z dimension. **(d)** Completely charged *a*-Se$_{30}$/Gr

cathode post potassium removal. The structure once again forms Se chains that are slightly condensed towards graphene surface.

## 2.2. Computational Details

All AIMD and DFT simulations were performed in Vienna Ab initio Simulation Package (VASP).[47] Inert core electrons were mimicked by Projector-augmented-wave (PAW) potentials and valance electrons were represented by plane-wave basis set with energy cut off at 550eV.[48, 49] The GGA-PBE exchange-correlation function was taken into account for all calculations.[50] AIMD simulations were run with 1 fs time interval, temperature set at 300 K within NVT ensemble, and 2 X 2 X 2 gamma centered k-meshes. For all DFT calculations, the conjugate gradient method was employed for energy minimization with Hellmann-Feynman forces less than 0.02 eV/Å and convergence tolerance set to $1.0 \times 10^{-4}$ eV. Gamma-centered 4 X 4 X 4 k-meshes were taken for good accuracy. Only for graphene-supported cathodes, GGA functional was inclusive of vdW correction to incorporate the effect of weak long-range van der Waals (vdW) forces. [51] All calculations for graphene-supported cathode were done with optPBE functional within vdW-DF-family. [52, 53]

## 3. Results and discussion

### 3.1. Potassium segregation at the interface

The presence of graphene substrate is reportedly beneficial for the Se electrode to control the dissolution of polyselenides in electrolytes and provide a supportive matrix to the volume fluctuating electrode.[39, 41, 46] However, graphene and pristine a-Se do not form a very reliable interface from a physicochemical perspective.[38] On a positive note, low interface strength (0.34

J/m$^2$) between Se and graphene indicates Se electrode interfaced with graphene can easily expand/contract during the battery cycle and evade high mechanical stresses.[37, 38] On the downside, the two materials are held by weak van der Waals (vdW) forces and have a very high potential gradient at the interface.[38] This condition is subjected to change as K atoms enter Se bulk due to prospective phase changes and different binding affinities of both atom types with graphene. Therefore, we expect distinct microstructural order in graphene supported a-K$_x$Se cathode from its free counterpart. Figure 3 presents relaxed atomic structures of a-K$_x$Se cathodes with and without graphene substrate, together with the respective variations of atomic K/Se ratio along the z-dimension.

To determine the influence of graphene substrate on the distribution of K and Se atoms, the atomic K/Se ratio profile is traced in three high K cathode configurations: a-K$_2$Se, a-K$_{1.6}$Se, and a-KSe. Figure 3(a) demonstrates the K/Se ratio profiles in cathode configurations with graphene support, while their graphene-free counterparts are analyzed in Figure 3(b). Simulation cells are divided into four bins (bin ID = 1, 2, 3, 4) along z-direction. In graphene-supported structures, the bottom 3.4 Å is not included in the bins considering it to be the vdW radius of graphene and represents volume occupied by only graphene. The rest of the simulation cell ( z - 3.4 Å) is divided into bins of width ranging from 3.18 Å to 3.8 Å depending upon the a-K$_x$Se thickness. For graphene-free a-K$_x$Se, bin widths ranged between 3.6 Å to 4 Å. The atomic K/Se ratios in each bin are marked with red and connected by blue line to view the pattern. The average K/Se ratio in the entire bulk (x in a-K$_x$Se ) is plotted as the dashed yellow line for comparison purposes.

In graphene-supported a-K$_x$Se (Figure 3(a)), two prominent regions can be noted based on K/Se ratio analysis: K-rich and K-deficient. K/Se ratios in bins 1 and 4 (close to graphene surface) clearly demonstrate higher K concentration (Figure 3(a)). In contrast, bin 2 (further from the graphene surface) has a low K concentration in all three cathode configurations (Figure 3(a)), i.e., K/Se = 2, 1.6, and 1. Combined K/Se ratio in bins 1 and 4 is continually above-average (yellow line) bulk K/Se ratio. A clear K concentration gradient is observed in sub-interfacial region (bin ID 1,4) and central region (bin ID 2,3). These plots indicate affinity of K atom to graphene surface and the possibility of K segregation at the interface. The balance of K concentration in electrode could be decided by observing bin-wise K/Se ratios (red) with respect to the average value (yellow line). The K concentration in completely discharged cathode a-$K_2$Se/Gr appears to be balanced. As shown in Figure 3(a1), bins 1 and 3 have average K/Se concentration value (*i.e.*, $x = 2$) that falls on the yellow line. However, K/Se concentration value in bins 2 and 4 is less and more than the average ($x=2$), respectively. In contrast, for a-$K_{1.6}$Se/Gr and a-KSe/Gr shown in Figure 3(a2-a3), K concentrations are higher than the average ($x=1.6$ and $x=1$, respectively) in three out of four bins (bin 1,3,4) suggesting a misbalance of K distribution in intermediate cathode structures.

The plots in Figure 3(b) represent K/Se ratios in graphene-free a-$K_x$Se cathode configurations and demonstrate alternative fluctuations of K/Se ratios within bins. It is important to remember that in these atomic representations of cathode without graphene, a-$K_x$Se cathodes are continuous periodic bulks. Understandably, a bin with high K concentration is followed by bin with low K concentration. No distinct pattern of K distribution can be recognized without a substrate. Moreover, K/Se ratio plot for low K concentration(K/Se <=1) cathode in Figure 3(b3) is similar to graphene supported one in Figure 3(a3), indicating that at low K content, segregation effect of graphene is reduced. Furthermore, all a-$K_x$Se structures in our study remained amorphous for both graphene-supported and graphene-free cases, as discussed in next section. It is unclear whether the segregation of K atoms at the graphene interface causes the formation of any new phase or phase boundary. The system size considered in our atomic study is too small to determine any phase transitions and phase boundaries. Nevertheless, the presence of graphene substrate creates a significant chemical gradient inside K-Se cathode intermediates (Figure 3(a)) and an imbalance in K concentration, which affects the site-specific energy of K atoms in the cathode[54] and stability of structural intermediates.

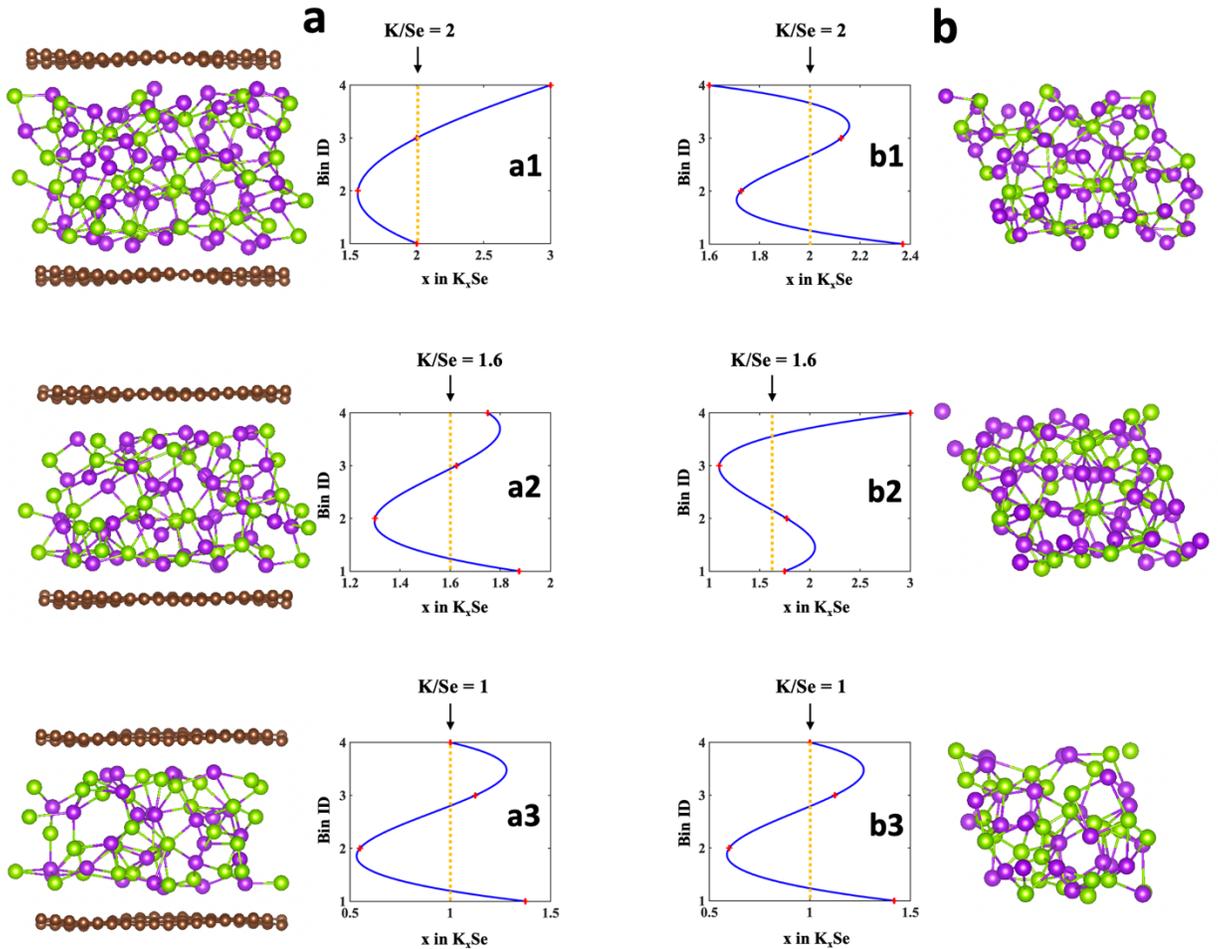

**Figure 3.** Degree of K segregation and K/Se ratio profile for **(a)** periodic a-$K_x$Se /Gr and **(b)** periodic a-$K_x$Se without any graphene support. Structures are divided into four bins along z dimension noted as Bin ID 1- 4. K/Se ratios in each bin are marked with red and connected by blue line to preview the pattern. Average K/Se ratio in the entire bulk is plotted as dashed yellow line

for comparison purpose. Graphene supported cathodes with over-all high average K ratio( >1) demonstrate higher K concentrations closer to graphene surface (Bin ID 1 and 3). In periodic a-$K_xSe$ cathodes without graphene substrate, K concentration peaked alternatively in bins. In cathodes with lower average K/Se ratio (<=1), distribution of K is nearly same irrespective of graphene presence.

### 3.2. Microstructural analysis

We observe in previous section that graphene presence causes K concentration gradient in the active $K_xSe$ cathode. Additionally, earlier reports [38] suggest that graphene interface can make the electronic states in a-Se more continuous and system more conductive. In reciprocal, the electronegative Se manipulates the surface chemistry of graphene substrate and activates it to provide additional ion storage sites [23, 55, 56]. In this section, we provide evidence that despite having a non-reactive surface, pristine graphene can enhance K-Se reactivity at the interface.

We first investigate short-range and long-range structural configuration of present systems by radial distribution function (RDF) analysis. The nearest neighboring distances between the atom pairs (Se-Se, Se-K, and K-K) in a-$K_xSe$ cathodes from RDF analysis are represented in Figure 4. Crystallinity is recognized by sharp singular distinct peaks in the plots, while broad peaks are characteristic of amorphous structure. RDF plots of completely discharged and partly discharged a-$K_xSe$ (x= 2 and 1) configurations in the graphene-supported systems (Figures 4(c-d)) are compared with graphene-free cases (Figure 4(a-b)). The important peaks for neighboring distances between Se-Se, Se-K, K-K atomic pairs are listed in Table 1. Crystalline c-$K_2Se$ and c-KSe systems are obtained from Materials Project database and their RDF analysis is used as a comparative standard to identify any signs of short-range crystalline order in amorphous atomic systems in our study.[57] In c-K$_2$Se (Figure S1(a)), nearest neighboring distance of 5.5Å for Se-Se atom pair denotes that no Se-Se covalent bonds are present in the structure. Similarly, no K-K bond pairs are present in the same structure. All Se atoms are bonded with K atoms with the bond length of 3.35Å in the crystal configuration shown in supporting information SI-1. In contrast, we observe more Se-Se bond pairs with bond length of 2.5Å in the atomic configuration of c-KSe with low K content (as shown in Figure S1(b)).

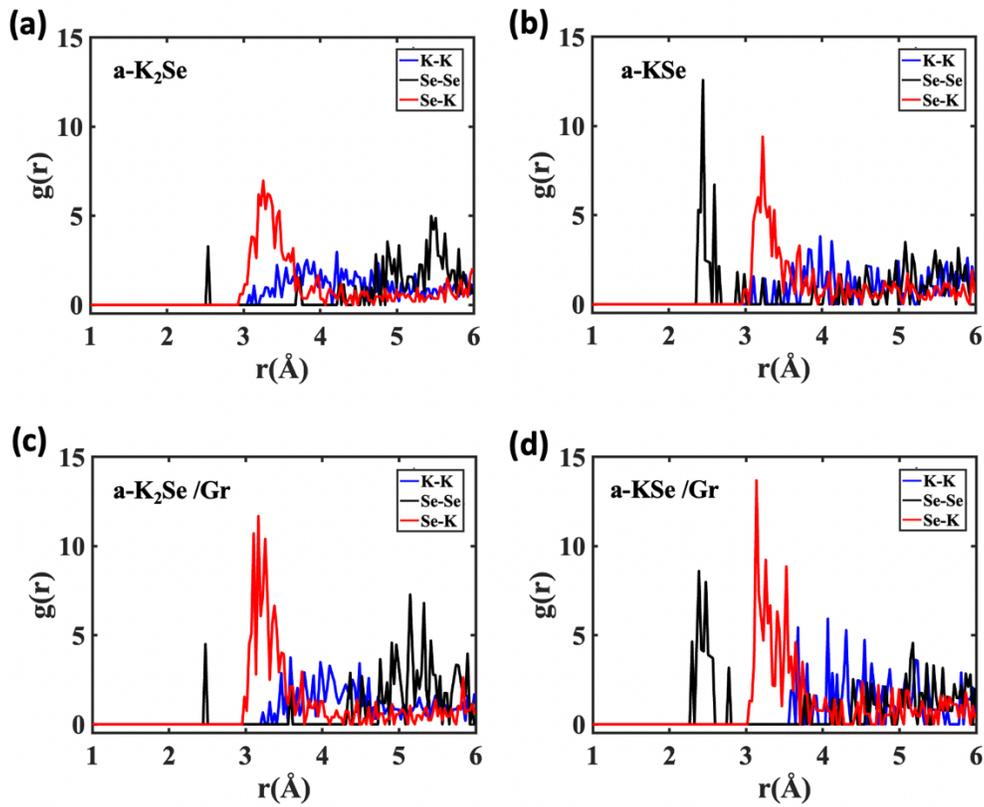

**Figure 4** Radial distribution function (RDF) plots exhibiting nearest neighboring distances between atomic pairs in graphene free configurations **(a)** a-K$_2$Se, **(b)** a-KSe, and graphene supported configurations **(c)** a-K$_2$Se/Gr, **(d)** a-KSe/Gr. Distances between atomic pairs are plotted as Se-Se with black, Se-K with red and K-K with blue.

**Table 1.** Nearest neighboring distances (Å) between Se-Se, Se-K and K-K in the $K_xSe$ cathodes with and without graphene substrate

| System | Se-Se | Se-K | K-K | Characteristic |
|---|---|---|---|---|
| a-$K_2Se$ | 2.5 Å, 4.9 Å, 5.5 Å | 3.25 Å | >3 Å | Amorphous |
| a-$K_2Se$/Gr | 2.4 Å, 4.4 Å, 5.2 Å | 3.2 Å | >3.2Å | Amorphous |
| c-$K_2Se$ | 5.5 Å | 3.35 Å | 3.8 Å | Crystalline |
| a-KSe | 2.5Å | 3.2Å | >3.0Å | Amorphous |
| a-KSe/Gr | 2.5 Å, 2.8 Å | 3.2Å | >3.5Å | Amorphous |
| c-KSe | 2.5Å, 3.7Å, 5.2Å | 3.3Å, 3.5Å | 4.3Å, 4.9Å, 5.4Å | Crystalline |

*Note: 'a-' denotes amorphous structure and 'c-' denote crystalline structures*

In graphene-free cathodes, the existence of long-range amorphous characteristics can be concluded from the RDF analysis. Peaks at 2.5 Å are noted in the RDF of a-$K_xSe$ systems, which indicate the presence of a few Se-Se covalent bonds. This peak is smaller and narrower in high K concentration cathode (a-$K_2Se$), while it is intense and broader in low K concentration cathode (a-KSe). As the concentration of K is twice the Se in a-$K_2Se$, Se-K bonds are prominent observations in Figure 4(a) with bond lengths of ~3.25Å and 3.35Å. The red peaks at 2.9-3.4Å are representative of Se-K distances and have a broad base. These peaks strongly contrast with very defined plots in crystalline configurations in Figure S1. In both a-$K_xSe$ systems, K forms strong covalent bonds with Se. Hence, no prominent K-K bond pairs are noticed in the RDF analysis. The K-K values in Table 1 represent the neighboring distances and not bond lengths. RDF plots for graphene-supported cathodes in Figure 4(c-d) show intense red peaks that are representative of Se-K bond. This increase in Se-K bond pairs along with evidence of K segregation at the interface in previous section presents a strong cue that graphene induces reactivity between Se and K in the system.

Though it remains unclear if the presence of graphene substrate induces any short-range crystallinity in the cathodes or not, graphene undoubtedly enhances K reactivity at the interface.

One standalone difference between high K cathode (a-$K_2Se$) and low K cathode (a-KSe) configurations in Figure 4 is the peak intensity at 2.5Å for Se-Se bonds. As earlier mentioned, a small peak at 2.5Å in a-$K_2Se$ (both graphene-free and graphene-supported cathodes) indicates the presence of only a few Se-Se covalent bonds. This peak becomes very prominent in a-KSe (both graphene-free and graphene-supported cathodes), signifying Se-Se bonds are more profound. Se retains their chained structures at low K concentrations. This causes key differences in the Se-K microstructures that are observed in low K (a-KSe/Gr) and high K (a-$K_2Se$/Gr) cathodes interfaced with graphene. In low K cathodes, the majority Se-Se bonds are intact, and interface contains potassium polyselenides with two to three Se atoms at the center surrounded by K atoms. When K concentration is further lowered (x < 1 in $K_xSe$), Se chains become longer. Meanwhile, as K concentration increases, most Se-Se bonds break to accommodate K. The resulting interface contains Se-K clusters with one Se atom at the center bound by multiple K atoms. Inspired by these inferences, we analyzed the Se-K clusters at graphene interface in a-KSe/Gr and a-$K_2Se$/Gr to determine their adsorption energy $E_{ad}$ over the graphene surface. Polyselenides in a-KSe/Gr (Figure 5(a)) are labeled as cluster-1 (Se-Se) and cluster-2 (Se-Se-Se). As discussed earlier, a-KSe/Gr was created after sequential depotassiation from a-$K_2Se$/Gr. Hence, Se-Se bonds in a-KSe/Gr are formed after K was removed (charging) and are not present due to the initial a-$Se_{30}$ chain structure. Meanwhile, a K saturation in a-$K_2Se$/Gr causes each Se atom to be surrounded by many K atoms (4 to 7). Typically, three Se-K clusters were present at the interface: $Se_1K_5$ labeled as cluster-3, $Se_1K_6$ labeled as cluster-4, and $Se_1K_7$ labeled as cluster-5 (Figure 5(b)). Only the Se atoms present near the graphene surface (bin ID 1) were bound to 6 or 7 K atoms, while Se in the central region (bin ID 2 and 3) were bound to 4 or 5 K atoms.

The surface adsorbed clusters were isolated from the bulk, and their adsorption energies (***E**ad*) over graphene substrate were determined as follows

$$E_{ad} = E_{total} - E_{cluster} - E_{graphene} \qquad (1)$$

Where ***E**total* is the energy of cluster over graphene substrate determined by DFT, ***E**cluster* and ***E**graphene* are the energy of the isolated cluster and pristine graphene substrate. Negative ***E**ad* denotes thermodynamically favored adsorption. We do not consider distinct translational or rotational configurations of Se-K clusters and limit our analysis to their existent orientation found in the parent bulk models (a-KSe/Gr and a-K$_2$Se/Gr). Stability of isolated clusters is realized from ***E**cluster* values (Figure 5(c)) in the order cluster-2 > cluster-1 > cluster-3 > cluster-4 > cluster-5. Polyselenides (cluster-1 and 2) are naturally more stable than high-K clusters (cluster 4 and 5). Our thermodynamic values indicate that high-K clusters cannot exist independently outside the graphene-supported cathode bulk. Therefore, high-K clusters prefer to strongly bind to graphene surface with highest binding energies marked in red in Figure 5(c) (-3.137 eV for cluster-4 and -3.419 eV for cluster-5). Interaction strength of the other three clusters (marked in blue in Figure 5(c)) with graphene is also reasonably high upon comparison with the literature.[58] Thus, Se-K reaction intermediates are less likely diffuse in electrolyte to cause shuttle effects due to their strong binding affinity with graphene substrate. Among interfacial clusters in low K cathode, cluster-2 (Se-Se-Se centered) is more stable and interacts more strongly with graphene substrate. As we observed an increase in binding energy with increase in Se content (Se-Se to Se-Se-Se), discharging may not be very favorable over graphene surface in low K cathode intermediates (with increased $Se_n$ chains).[58, 59]

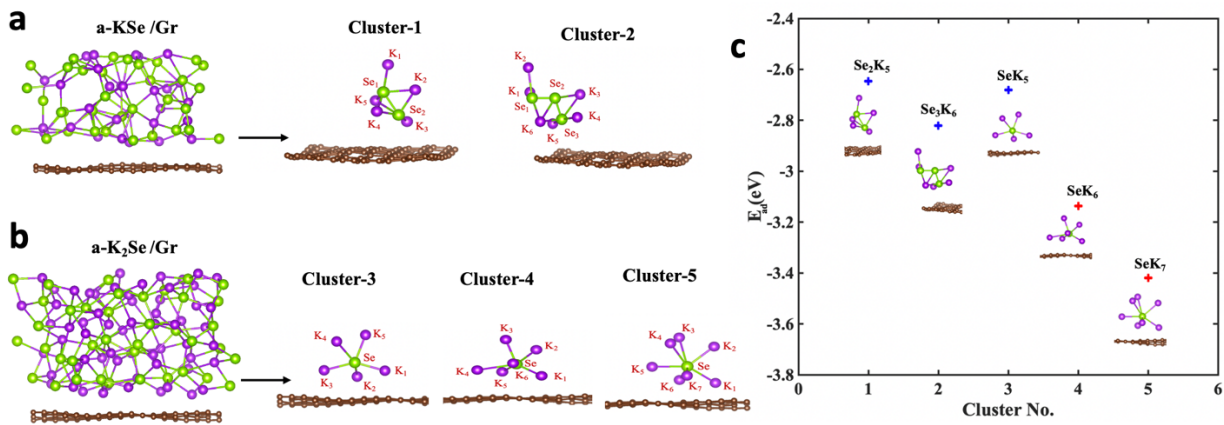

**Figure 5. (a)** Side view of relaxed a-KSe/Gr structure. Majority Se-Se bonds are intact, and interface contains potassium polyselenides with two to three Se atom chains at the center. These clusters are numbered as 1, 2. **(b)** Side view of relaxed a-$K_2$Se/Gr structure. Majority Se-Se bonds have broken to accommodate K and interface contains Se-K clusters with one Se atom at the center surrounded by K atoms. Clusters are numbered as 3,4 and 5. **(c)** Binding energies $E_{ad}$ of different K-Se clusters noted at the interface with graphene substrate. $E_{ad}$ of cluster 4 and 5 which are least stable in isolated state but bind strongly with graphene are marked in red.

Our previous study on physiochemical analysis of Se-Gr interface shows that loose Se ($Se_1$) atoms in a-Se in the interfacial region exhibit individualism and adsorb on graphene surface by gaining electrons leading to p-type doping in the latter. [38] Once Se chains dissociate at high K concentrations, K starts to accumulate in the interfacial region because of their strong binding affinity with graphene.[60] Presence of K in interfacial region counteracts the doping characteristics of graphene. Density of states (DOS) analysis (see Supporting Information SI-2) exhibits a negative shift in Dirac cone of graphene in the presence of high K clusters, indicating n-type doping characteristics. In the presence of high K near the interface, graphene gains more electrons and presents electron rich surface that acts as an electrocatalyst for further Se-K reactivity in the interfacial region. However, these effects could be highly regional on graphene surface leading to irregular K distribution. This is because mobility of charge carriers has been previously shown to decrease with increase in K surface dopants. [61]

### 3.3. Electrochemical Mechanism

Operation of KIB is based on rocking-chair principle of LIBs, as K ions shuttle between anode and cathode through an electrolyte. To navigate this K ion shuttling, a chemical potential difference must exist between cathode ($\mu^{cathode}$) and anode ($\mu^{anode}$), which is referred to as open circuit voltage (*OCV*).

$$OCV = -\frac{\mu_K^{cathode} - \mu_K^{anode}}{zF} \qquad (2)$$

Here, **F** is Faraday's constant, and **z** is the electronic charge transported by K in the electrolyte (*z* = 1 for K in non-conducting electrolyte). To theoretically calculate **OCV** for $K_xSe$ cathode, K metal anode is considered with a constant chemical potential equivalent to K metal's Gibbs free energy (depicted in Figure 6(a)). Thus, electrical energy[62] gained in discharging between $K_ySe$ and $K_xSe$ (x > y) is given by difference in Gibbs free energy (**G**) of the two compounds as

$$E = -[G_{K_xSe} - G_{K_ySe} - (x-y)G_K] \qquad (3)$$

$$E = -\Delta G \tag{4}$$

Where $G_K$ is the total Gibbs free energy of a single K atom in metallic K unit cell, and *(x - y)* represents K atoms intercalated in cathode during discharging. This leads us to calculate average intercalation voltage in cathode between two intercalation limits as

$$V = \frac{E}{(x-y)F} \tag{5}$$

Gibbs free energies calculated in our study are in electronvolts (eV) and are plotted for intermediate a-$K_x$Se/Gr structures (0 <= x < 2) in Figure 6(b). Therefore *F* is neglected in the above equation.[63] Average voltage profiles between final discharged cathode a-$K_2$Se and cathode intermediates ( a-$K_x$Se with 0 <= x < 2) as a function of K content are shown in Figure 6(c). A sloping voltage curve in range 1.4 – 0.38 V (pink curve in Figure 6(c)) is the result of amorphous Se forming a solid solution with K in absence of any substrate. The voltage plot slopes downward with a small spike at x=1 ( KSe ). It is suggestive of two-step reaction process where $K_2Se_2$ (KSe) is initially formed before being further reduced to $K_2$Se, the final discharged product. We must mention here that different reported reduction pathways exist for Se in K-Se batteries based on electrolytes and Se microstructures. [64, 65] $Se_n$ chains undergo a single-step conversion to $K_2$Se in carbonate-based electrolytes, which can sometimes become two-step conversion due to low reactivity (Se → $K_2Se_2$ →$K_2$Se).[30] Alternatively, several intermediate $K_x$Se are formed during $Se_n$ reduction when ether-based electrolytes are used. [64] Since electrolytic influences are not taken into consideration in the present study, we can safely assume that electrochemical mechanisms observed here are purely driven by Se molecular structure. As previously established,

Se cathode cannot successfully operate without any host matrix. Capacity of pristine Se in KIB has been previously shown to drop to zero after first electrochemical cycle regardless of using carbonate-based electrolytes.[25] Thus, high theoretical capacity of Se cathode can only be leveraged by combining Se with a C-based matrix.[65]

In comparison to porous C, hexagonal C lattice-based matrix enclosing Se cathodes exhibit better performances for LIB.[41, 46, 66, 67] This is partly due to high conductivity of hexagonal C lattice and partly due to inability of Se to form covalent interactions with lattice surface, thereby preventing the loss of active Se electrode.[38] Moreover, host matrix can also have significant impact on electrochemical reactions in Se cathode. This concept is less understood in literature mostly due to complex design of Se-based composite cathodes. The blue plot in Figure 6(c) shows intercalation voltage profile of a-$K_x$Se cathode supported by hexagonal C lattice, i.e., graphene in this case. Major inferences that can be drawn from the plot are: First, the discharge voltage for intercalation limit x = 0 and x = 2 is 1.55 V. Second, multiple reaction products may be present in the system due to incomplete reaction of Se.

Two high voltage peaks are noted at x = 0.8 and x=1.6 in blue plot of Figure 6(c). These peaks correspond to high energy structures *ii* and *v*, as indicated in Figure 6(b). The energy of a-Se/Gr cathode dips during the discharging process except near the intermediates *ii* and *v* where the sudden energy spikes are noted. The plot indicates a difficult electrochemical reduction of Se cathode with possible formation of multiple reaction products. From atomistic point of view, the rise in energy between intermediates *i* and *ii* is due to the stability of long Se chains on the graphene surface (Se-Se-Se vs. Se-Se in Figure 6(c)) that causes an inconvenience in discharging. The higher binding affinity between graphene and polyselenides with increased Se content (Se-Se to Se-Se-Se) could cause difficulty in Se chain breakage to store more K. Post this stage (x=0.8), Se continues to react with incoming K until high K concentration (x =1.6) is reached corresponding to peak *v*. We know from our discussion in Section 3.1 and 3.2 that graphene is causing strong K segregation at the interface in high K intermediates and impacting the electrochemical reaction by inducing new reactive sites near the interface. This causes irregular distribution of K through the cathode resulting in incomplete Se reduction at x=1.6 ($K_{1.6}Se$). Similar peaks (like in Figure 6(b)) are reported in previous studies where electrode was recognized to constitute two distinct phases.[68, 69] While the detailed microstructural and electrochemical analysis of graphene supported Se-K cathode strongly suggest possibility of multiple phases in the system, it is challenging to draw conclusions from nano-scale system. Potential interface-induced phase changes in these heterostructure electrodes could be further explored by multiscale modeling. Upon further discharging (K intercalation), all Se present in the system reacts with K to form a thermodynamically stable discharged product $K_2Se$. We deduce that graphene has modulating favorability towards discharging of Se cathode due to its surface characteristics and differential preferences for $K_xSe$ intermediate systems.

From Figure 6(c), we draw that operating a-Se/Gr cathode at high voltage conditions (~2V) can lead to the formation of cathode intermediates (ii and *v*) which represent thermodynamical energy barriers in the process of intercalation/deintercalation and could cause irreversible capacity losses. If we observe the intercalation voltage profile in Figure 6(c) upon ignoring these thermodynamical barriers (dashed red plot), the voltage remains in 1.55 - 1.38 V range and exhibits a plateau-like profile. It is possible for Se to undergo a single-step reaction with K to form $K_2Se$ if the applied voltage is 1.55 V, much like in the case of Li-Se.[70]

$Se + 2K^+ + 2e^{-1} \rightarrow K_2Se$            *(6)*

This voltage range for electrochemical reaction is also close to cathode voltage reported in a study with Se hosted by carbon nanotube anchored microporous C.[27] Our results feature strong dependence of Se-K electrochemistry on the interface presented by graphene matrix in nanostructured electrodes.

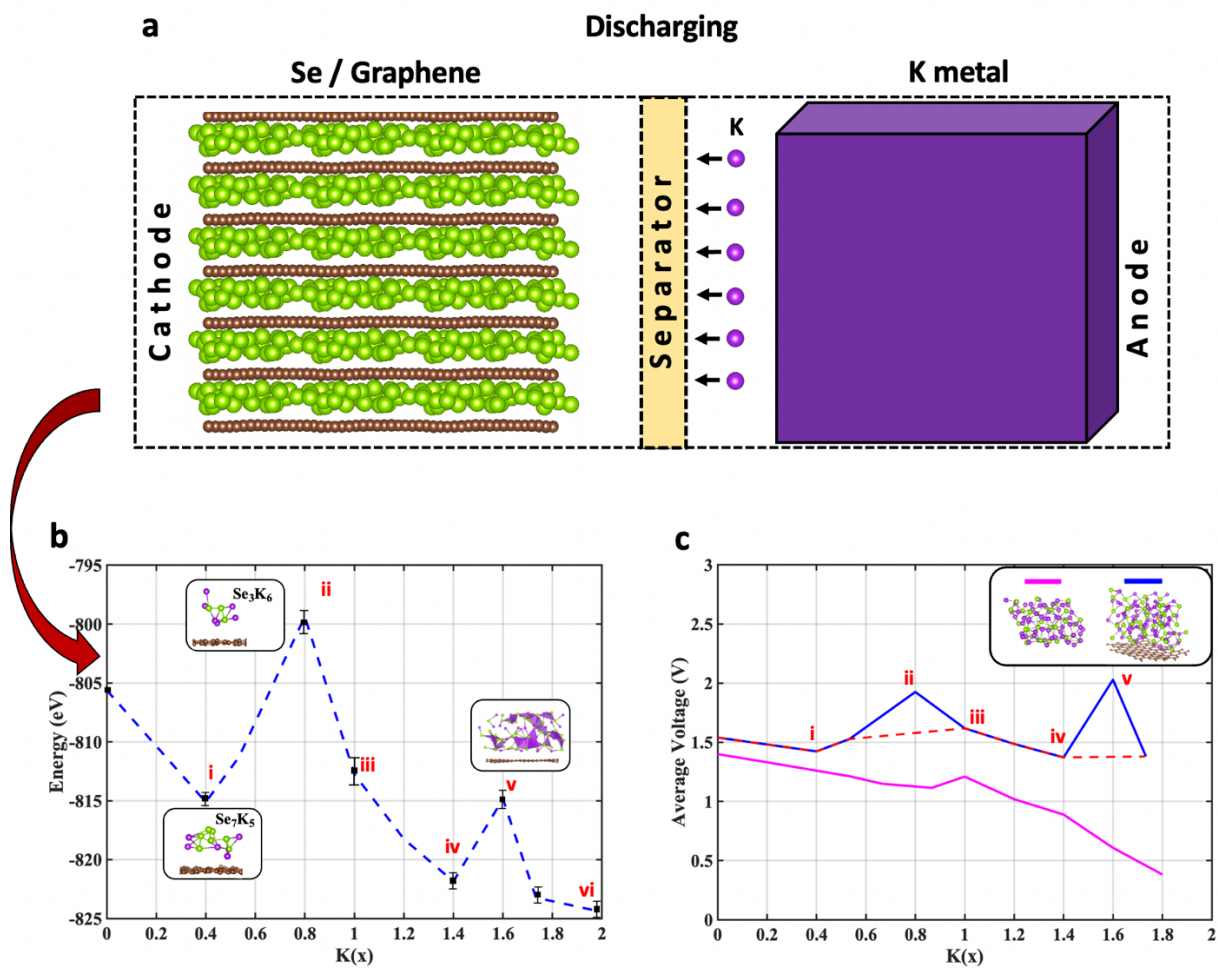

**Figure 6. (a)** Selenium-graphene heterostructure cathode half-cell during discharging process in KIB. **(b)** Energy of relaxed graphene supported a-$K_xSe$ (0 <= x < 2) cathode intermediates labelled *i* to *vi*. **(c)** Average discharge voltage of a-$K_xSe$ cathode intermediates with $K_2Se$ as final

discharged product. The voltage profiles of Se-K alloying cathode with and without graphene substrate are plotted in blue and pink, respectively.

## 4. Conclusions

To conclude, we modeled graphene-supported $K_xSe$ cathode identical to experimentally designed electrodes and investigated the effect of graphene interface on electrochemical mechanism by AIMD and DFT. Our results are first to highlight strong dependence of electrochemical mechanism in Se cathode on host structure. The key findings presented in the study are:

- Graphene substrate causes a substantial chemical gradient inside the K-Se cathode by inducing enhanced Se-K reactivity in the interfacial region.
- Surface chemistry of graphene interfaced with $K_xSe$ cathode modulates based on its differential preference for Se and K atoms. At low K concentrations, graphene causes difficult K intercalation (discharging) because of its strong binding preference with longer $Se_n$ chains ($E_{ad}$ = -2.646 eV for n=2 and $E_{ad}$ = -2.82 eV for n=3). Meanwhile at high K concentrations, K density near interface increases causing the formation of Se-K clusters with a high K atom count ($SeK_6$, $SeK_7$). These clusters are not stable without substrate and interact strongly with graphene surface through binding energies ($E_{ad}$) as high as -3.137 eV for $SeK_6$ and -3.419 eV for $SeK_7$.
- This modulating favorability of graphene towards discharging of Se cathode could cause formation of reaction intermediate with thermodynamically challenging K insertion and extraction.

- To avoid irreversible capacity losses, graphene-supported Se cathode should operate in the voltage range of 1.55V to 1.38V, which will lead to a single step reaction near 1.55 V with $K_2Se$ as discharged product.

**Supporting Information**

Radial distribution function (RDF) of crystalline $K_xSe$; Density of States (DOS) analysis of high K clusters with graphene surface

**Acknowledgements**

V.S. and D.D. were supported by NSF (Award Number -1911900 and 2237990) during the study. D.D. acknowledges the Extreme Science and Engineering Discovery Environment (XSEDE) for the computational facilities (Award Number − DMR180013).

**Competing Interest**

The authors declare no competing financial interest

**Data Availability**

The data reported in this paper is available from the corresponding author upon reasonable request.

**Code Availability**

The pre- and postprocessing codes used in this paper are available from the corresponding author upon reasonable request. Restrictions apply to the availability of the simulation codes, which were used under license for this study.

## Author Contributions

V.S. contributed to the work with conception of the project, computation, and manuscript preparation. D.D. discussed results with V.S. and contributed to the manuscript preparation. All authors have given approval to the final version of the manuscript.

## Author Information

**Vidushi Sharma** is currently working as a Staff Research Scientist at IBM Almaden Research Center in San Jose, CA, USA where she joined in November, 2021 after completed her Ph.D. from New Jersey Institute of Technology (NJIT) in Mechanical Engineering.

**Dibakar Datta** is currently an Associate Professor at the Department of Mechanical and Industrial Engineering, Newark College of Engineering, New Jersey Institute of Technology (NJIT), Newark, USA. He finished his Ph.D. at Brown University.

# Table of Contents (TOC)

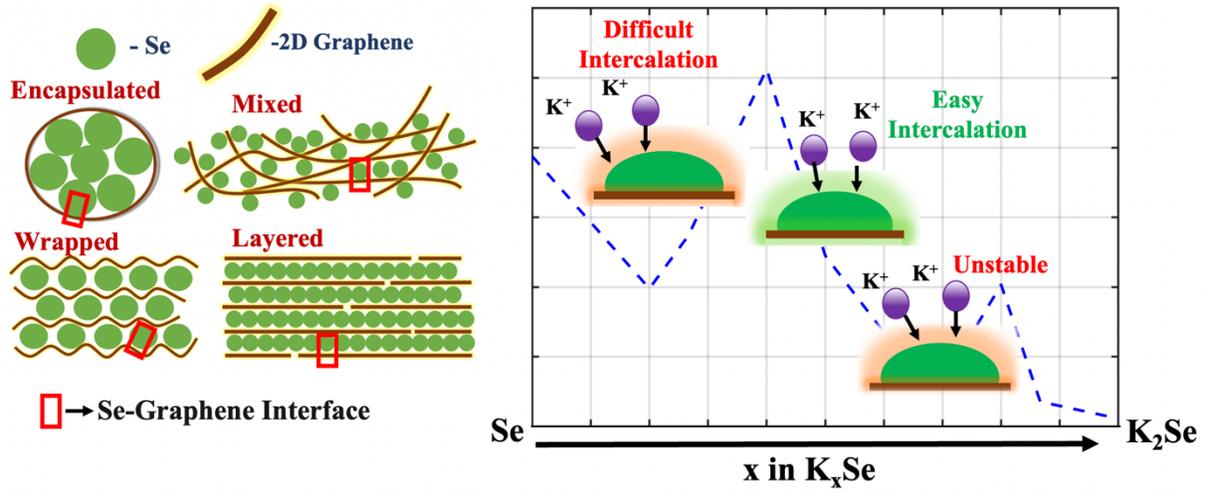